# Direct observation of quasi-particle band in CeIrIn$_5$: Angle-resolved photoemission spectroscopy study

Shin-ichi Fujimori[1], Atsushi Fujimori[1,2], Kenya Shimada[3], Takamasa Narimura[4], Kenichi Kobayashi[4], Hirofumi Namatame[3], Masaki Taniguchi[3], Hisatomo Harima[5], Hiroaki Shishido[6], Shugo Ikeda[7], Dai Aoki[8], Yoshifumi Tokiwa[9], Yoshinori Haga[7], and Yoshichika Ōnuki[6,7]

[1]*Synchrotron Radiation Research Unit, Japan Atomic Energy Agency, SPring-8, Sayo, Hyogo 679-5148, Japan*
[2]*Department of Complexity Science and Engineering, University of Tokyo, Kashiwa, Chiba 277-8561, Japan*
[3]*Hiroshima Synchrotron Radiation Center, Hiroshima University, Higashi-Hiroshima, Hiroshima 739-8526, Japan*
[4]*Graduate School of Science, Hiroshima University, Higashi-Hiroshima, Hiroshima 739-8526, Japan*
[5]*Department of Physics, Kobe University, Nada, Kobe 657-8501, Japan*
[6]*Graduate School of Science, Osaka University, Toyonaka, Osaka 560-0043, Japan*
[7]*Advanced Science Research Center, Japan Atomic Energy Agency, Tokai, Ibaraki 319-1195, Japan*
[8]*Institute for Material Research, Tohoku University, Sendai 980-8578, Japan*
[9]*Los Alamos National Laboratory, Condensed Matter and Thermal Physics, MST-10, MS K764, Los Alamos, New Mexico 87545, USA*
(Dated: February 13, 2006)

We have performed a high-resolution angle resolved Ce $4d$-$4f$ resonant photoemission experiment on the heavy fermion superconductor CeIrIn$_5$. We have observed a quasi-particle band which has an energy dispersion of $\sim 30$ meV in the Ce $4f$ on-resonance spectra. The result suggests that although the $4f$ spectra are dominated by the localized/correlated character, the small itinerant component is responsible for the superconductivity in this compound.

## I. INTRODUCTION

The discovery of superconductivity in heavy fermion (HF) $f$-electron compounds near a magnetic quantum critical point (QCP) has stimulated the studies of superconductivity in this class of materials [1, 2]. In these compounds, both magnetism and superconductivity are originated from the $f$ electrons, which are considered to have dual character, i.e., both localized and delocalized character [3]. The recently discovered Ce$T$In$_5$ ($T$=Rh, Ir, Co) compounds have attracted much attention because they are antiferromagnets or superconductors depending on the chemical composition and pressure, which suggests that the system is located near a QCP. Therefore they are good target materials to study how magnetism and superconductivity are related with each other. At ambient pressure, CeRhIn$_5$ [4] becomes antiferromagnetic below $T_N = 3.8$ K, while CeIrIn$_5$ [5] and CeCoIn$_5$ [6] become superconducting below $T_C = 0.4$ K and 2.3 K, respectively. For the CeRh$_x$Ir$_{1-x}$In$_5$ alloy compounds, it was reported that the antiferromagnetic (AF) long-range order and the superconductivity coexist for $0.3 < x < 0.6$ [7]. These differences in the ground state properties of the different Ce$T$In$_5$ compounds are thought to be originated from the differences in the lattice constants of the quasi-two-dimensional (2D) CeIn$_3$ layers, and hence the difference in the $f$-$p$ hybridization strength. Therefore, this system provides an unique opportunity to study the relationship between magnetism and superconductivity in $f$-electron systems.

There are several experimental studies on the electronic structure of $f$ states in these compounds. For antiferromagnetic CeRhIn$_5$, Shishido et al. [8] have performed de Haas van Alphen (dHvA) experiments, and found that the obtained branches are essentially the same as those of LaRhIn$_5$, suggesting that the Ce $4f$ electrons in CeRhIn$_5$ are localized. The localized nature of Ce $4f$ electrons in this compound is also evident from the dHvA experiments on Ce$_x$La$_{1-x}$RhIn$_5$ by Los Alamos group [9] and the optical conductivity measurements[10]. For the superconducting CeIrIn$_5$, on the other hand, there has been a controversy between the localized and delocalized pictures for the Ce $4f$ electrons. A nuclear quadrupole resonance (NQR) study has suggested that the Ce $4f$ electrons in this compound are much more itinerant than known other Ce HF based compounds [11]. dHvA measurements on this compound have also suggested that the observed branches are well explained by the itinerant $4f$-electron model [8, 12]. In addition, Harrison et al. [9] have compared the dHvA branches of CeIrIn$_5$ and LaIrIn$_5$, and found that the expansion of Fermi surface (FS) volume on going from LaIrIn$_5$ to CeIrIn$_5$. This implies the formation of quasi-particle band in CeIrIn$_5$. All these results suggest that $4f$ electrons in CeIrIn$_5$ have itinerant character. On the other hand, photoemission studies have been made for this compound, and the localized character of $f$-electrons has been observed. In a previous paper, we have studied Ce$T$In$_5$ ($T$=Rh, Ir) by angle-resolved photoemission spectroscopy (ARPES) using low energy photon ($h\nu$=21.2 eV) and Ce $3d$-$4f$ resonant photoemission (RPES) [13]. The RPES spectra of both compounds were understood within the framework of single impurity Anderson model (SIAM), suggesting that the nearly localized nature of Ce $4f$ states in these compounds. Since we could not observed delocalized spectral features of the Ce $4f$ states in CeIrIn$_5$, we concluded that the Ce $4f$ electrons in CeRhIn$_5$ and CeIrIn$_5$ were nearly localized, but they were slightly delocalized in CeIrIn$_5$ compared with CeRhIn$_5$. Because the spectral difference between CeIrIn$_5$ and CeRhIn$_5$ was very small, the origin of the large difference in their ground state properties was not well understood from the photoemission spectra.

To probe the localized/delocalized nature of $f$-states, ARPES experiment is the most powerful experimental method since the observation of dispersive $f$-bands is the direct evidence for the itinerant character of $f$ states. There have been some attempts to detect the energy dispersions of the Ce $4f$ band by ARPES. Arko et al. [14] have reported an ARPES study of CeSb$_2$, and observed the intensity modulation of the Ce $4f$ peak near the Fermi level ($E_F$) as a function of momentum. However, they measured the spectra at $h\nu = 60$ eV, where contributions not only from Ce $f$ but also Ce $d$ and Sb $s, p$ states were substantial, and therefore it was difficult to extract the Ce $4f$ contribution unambiguously [15]. To distinguish the Ce $4f$ contribution from the other states, the RPES technique has been utilized. Denlinger et al. [16] measured angle-resoled RPES (AR-RPES) spectra of CeRu$_2$Si$_2$, and found the amplitude variation of the Ce $4f$ derived component near $E_F$. However, they could not observed the experimental energy dispersion of the Ce $4f$ derived component, probably due to the weakness of the dispersion compared to the energy resolution ($\Delta E \sim 60$ meV).

In the present paper, we report on a high-resolution ($\sim 30$ meV) AR-RPES study of CeIrIn$_5$ in the $4d$-$4f$ excitation region to study the dispersion of the $4f$ band, thereby clarifying the $f$-electron states in CeIrIn$_5$. The improved energy resolution has enabled us to detect the energy dispersion of the Ce $4f$ states in CeIrIn$_5$ as we shall discuss below.

## II. EXPERIMENTAL

The experiment was performed at the linear undulator beamline (BL-1) of a 700 MeV electron-storage ring (HiSOR) in Hiroshima Synchrotron Radiation Center (HSRC), Hiroshima University. The photon energies of $h\nu \sim 122$ eV and 112 eV were used for $4d$-$4f$ on- and off-resonance, respectively. The energy resolution of 18 meV and 30 meV were used for the angle-integrated and angle-resolved photoemission experiments, respectively. Single crystals of CeRhIn$_5$ and CeIrIn$_5$ were grown by the self-flux method described in Ref. [4]. Clean surfaces were obtained by in situ cleaving the samples parallel to the $a$-$b$ plane. The sample temperature was kept at 10 K during the course of the measurements. The position of $E_F$ was carefully determined using evaporated gold film.

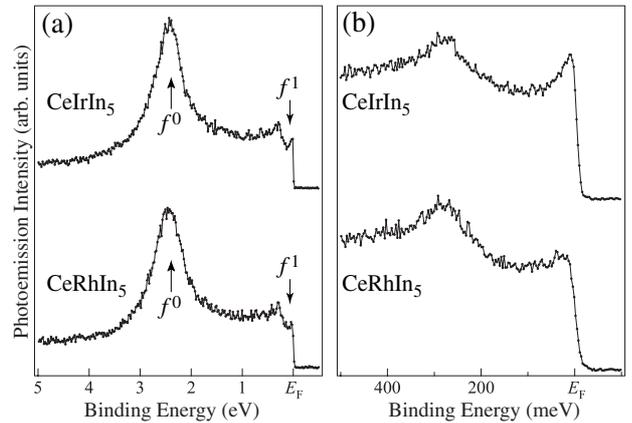

FIG. 1: $4d$-$4f$ resonant photoemission spectra of CeIrIn$_5$ and CeRhIn$_5$ ($h\nu = 122$ eV). (a)Wide range. (b)Near $E_F$ part.

## III. RESULTS AND DISCUSSION

Figure 1(a) shows angle-integrated $4d$-$4f$ RPES spectra of CeRhIn$_5$ and CeIrIn$_5$. The spectra show an intense peak at around 2.5 eV below $E_F$ corresponding to the $f^0$ final state, and weak structures near $E_F$ corresponding to the $f^1$ final states. The $4f^1$ final-state feature shows peak at $\sim 280$ meV arising from the spin-orbit interaction. These spectral features have been observed in other Ce-based compounds, and understood within the framework of the SIAM as in the case of $3d$-$4f$ RPES study for these compounds [13]. The intensity of the $f^1$ final-state signal near $E_F$ is much weaker than that of the $f^0$ final state peak located at $\sim 2.5$ eV. It has been pointed out that the Ce $4f$ electrons in the surface layers are more localized than in the bulk electronic structure [17]. Sekiyama et al. [18] have demonstrated that the surface-sensitive $4d$-$4f$ RPES spectra of HF Ce-based compounds are significantly different from the bulk-sensitive $3d$-$4f$ RPES spectra. In fact, the $f^0$ peak intensity is much stronger in the present $4d$-$4f$ RPES spectra than in the previous $3d$-$4f$ RPES spectra [13], indicating that Ce $4f$ electrons in the surface layers are more localized than in the bulk.

Nevertheless, the weakness of the $4f^1$ signal compared with other itinerant $f$-electron compounds in the present spectra are consistent with the nearly localized character of $4f$ electrons in the bulk. To address the question of whether the present $4d$-$4f$ RPES can be used for probing the bulk electronic structure of these compounds, we note that because the Ce $4f$ electrons in these compounds are nearly localized, the Ce $4f$ electrons in the surface layer must be almost completely localized. This means that the surface component of the Ce $4f$ states should have very small spectral weight in the delocalized ($f^1$) part of the spectra. Therefore, we assume that the $f^1$ peak in these spectra can be mostly considered largely the bulk origin. Figure 1(b) shows the near $E_F$ part of the RPES spectra. The spectra consist of the $f_{5/2}{}^1$ final state peak located at $E_F$ (in fact the tail of the Kondo peak located above $E_F$) and the $f_{7/2}{}^1$ peak located $\sim 280$ meV below $E_F$. The clearest difference between the spectra of CeIrIn$_5$ and CeRhIn$_5$ is that the spectral intensity at $E_F$, the Ce $4f_{5/2}$ peak or the tail of the Kondo peak, is higher in CeIrIn$_5$ than in CeRhIn$_5$. Because its intensity is higher for more strongly hybridized compound, the absence of the sharp peak structure in CeRhIn$_5$ suggests that Ce $4f$ electrons in this compound is almost completely localized in CeRhIn$_5$ while those in CeIrIn$_5$ have some itinerant character. These results are consistent with the results of optical conductivity measurements where the hybridization gap is observed in CeIrIn$_5$ but not in CeRhIn$_5$ [10].

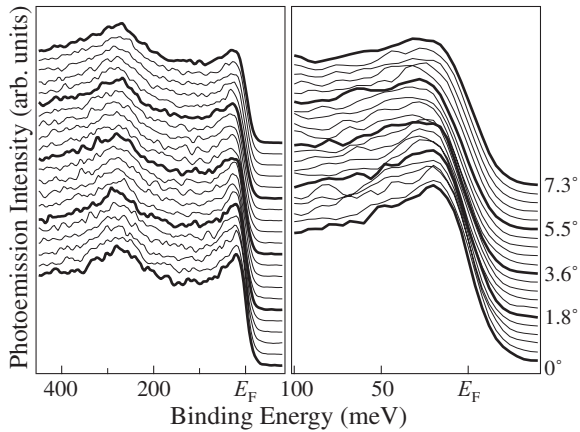

FIG. 2: On-resonance ARPES spectra near $E_F$, taken at $h\nu=122$ eV.

We have further performed AR-RPES experiments on CeIrIn$_5$ to reveal its momentum dependence of the Ce $4f$ spectrum. Figure 2 shows the angle-resolved $4d$-$4f$ RPES spectra of CeIrIn$_5$. The left panel shows the spectra in a wide energy range, and the right panel does details of the spectra near $E_F$. The emission angle relative to the surface normal is also presented. It can be seen that the spectra show a clear angular dependence. In particular, the intensity and the shape of $4f_{5/2}$ peak located just below $E_F$ show a clear momentum dependence. Its intensity is largest around $\theta = 0°$, and becomes weaker when the emission angle is increased. Moreover, the peak shows a small but distinct energy dispersion. The peak moves toward the deeper binding energy side as it approaches $\theta \sim 4°$, and it again moves toward $E_F$ as the emission angle is further increased. These changes are observed only in the $4f_{5/2}$ final state peak, and are not seen in the $4f_{7/2}$ final state peak. This suggests that the changes are not caused by matrix-element effects but by band dispersion.

Here, we consider the position of the present ARPES cut in $\bm{k}$ space. The momentum of photoelectron parallel and perpendicular to the surface are given by

$$k_\parallel = \sqrt{\frac{2m}{\hbar^2} E_{\text{kin}}} \sin\theta, \qquad (1)$$

$$k_\perp = \sqrt{\frac{2m}{\hbar^2}(E_{\text{kin}}\cos^2\theta + V_0)}, \qquad (2)$$

where $E_{\text{kin}}$ is the kinetic energy of the photoelectron, $V_0$ is the inner potential, and $\theta$ is the emission angle of the photoelectron relative to the surface normal. To determine the $k_\perp$ value, we need to determine the inner potential $V_0$. In the previous ARPES experiments on CeIrIn$_5$ [19], the $V_0$ is estimated to be 15 eV. However, this value is slightly over estimated because they do not include the contributions from the momentum of incident photon, which is not negligible in this photon energy range. In fact, this value is somewhat larger than those of other HF compounds such as CeRu$_2$Si$_2$ (12 eV) [16] and CeNi$_2$Ge$_2$ (13.8 eV) [20]. Therefore, we assume that the inner potential is 12 eV, which can be regarded as a typical value for HF materials. Figure 3(b) shows the position of the APRES cut as a function of $k_\perp$ and $k_{//}$ for on-resonance ARPES spectra. Because the escape depth of the photoelectron in the present experiment is about $5-10$ Å, the momentum broadening for the $k_\perp$ direction should be $\Delta k_\perp \sim 0.1 - 0.2$ Å$^{-1}$. This broadening width ($\Delta k_\perp = \pm 0.1$ Å$^{-1}$) is also shown as shaded area in Fig. 3(b). Since the contributions from the high symmetry lines are dominant in ARPES spectra, we assume that the present AR-RPES spectra probe the high-symmetry Γ-X line of the 8th Brillouin zone. The spectrum measured at $\theta = 0°$ corresponds to the Γ point, and that measured at $\theta = 7°$ does the X point.

Figure 3(c) shows the density plot of ARPES spectra. The momentum dependences of the ARPES spectra are summarized as follows. The $4f_{5/2}$ peak located at around $E_F$ show small energy dispersions. On the other hand, for the $4f_{7/2}$ peak, while their positions are almost unchanged, their widths have some momentum dependences. Therefore, they have very different behaviors, suggesting again that these changes are not due to the matrix element effects. To see the angular dependent behavior of the peak near $E_F$ more clearly, we have divided the AR-RPES spectra by the Fermi-Dirac function convoluted with a Gaussian function representing the instrumental resolution (30 meV). This method has been utilized to remove the effect of the Fermi cut-off and to reveal spectral feature around $E_F$. Figure 3(d) shows the density plot of the ARPES spectra in the $E-k$ plane divided by the Fermi-Dirac function. In this plot, we have marked the positions of the peak maximum as solid circles as a function of binding energy and momentum. In the near $E_F$ part of the spectra, a dispersive feature is clearly recognized. The size of this energy dispersion is estimated to be about 30 meV. Mena et al. [10] have measured the optical conductivity of Ce$M$In$_5$ ($M$=Rh, Ir, and Co), and observed a hybridization gap of $\Delta = 50 - 70$meV for CeIrIn$_5$. This energy scale is of the same order as the energy dispersion observed in the ARPES spectra, and consistent with our observation of the narrow hybridized band. This behavior of narrow $f$-band can be explained within the framework of the renormalized $f$-band model[21], according to which the renormalized $f$ band is mixed with the strongly dispersive non-$f$ band, and the weakly dispersive hybridized bands are formed near $E_F$. In the figure, we have plotted the peak positions of ARPES spectra measured by He I ($h\nu$=21.2 eV) [13] as open circles, which can be regarded as the contributions mostly from In 4$p$-bands. Although these spectra do not correspond to the same position of the ARPES scan measured by $h\nu$=122 eV for $k_\perp$ direction, it is known that ARPES spectra with He I represent the contributions mostly from the high symmetry lines [22]. Therefore, we take these ARPES spectra as the contributions from the non-$f$ bands in high symmetry Γ-X line. The high intensity part in the quasi-particle bands is distributed inside of Fermi surface crossing points of the non-$f$ bands, suggesting that the features can be understood within the framework of renormalized band theory. Therefore, in the present case, the narrow 4$f$ quasi-particle band is originated from the hybridization between the renormalized Ce 4$f$ states and the In 4$p$ states.

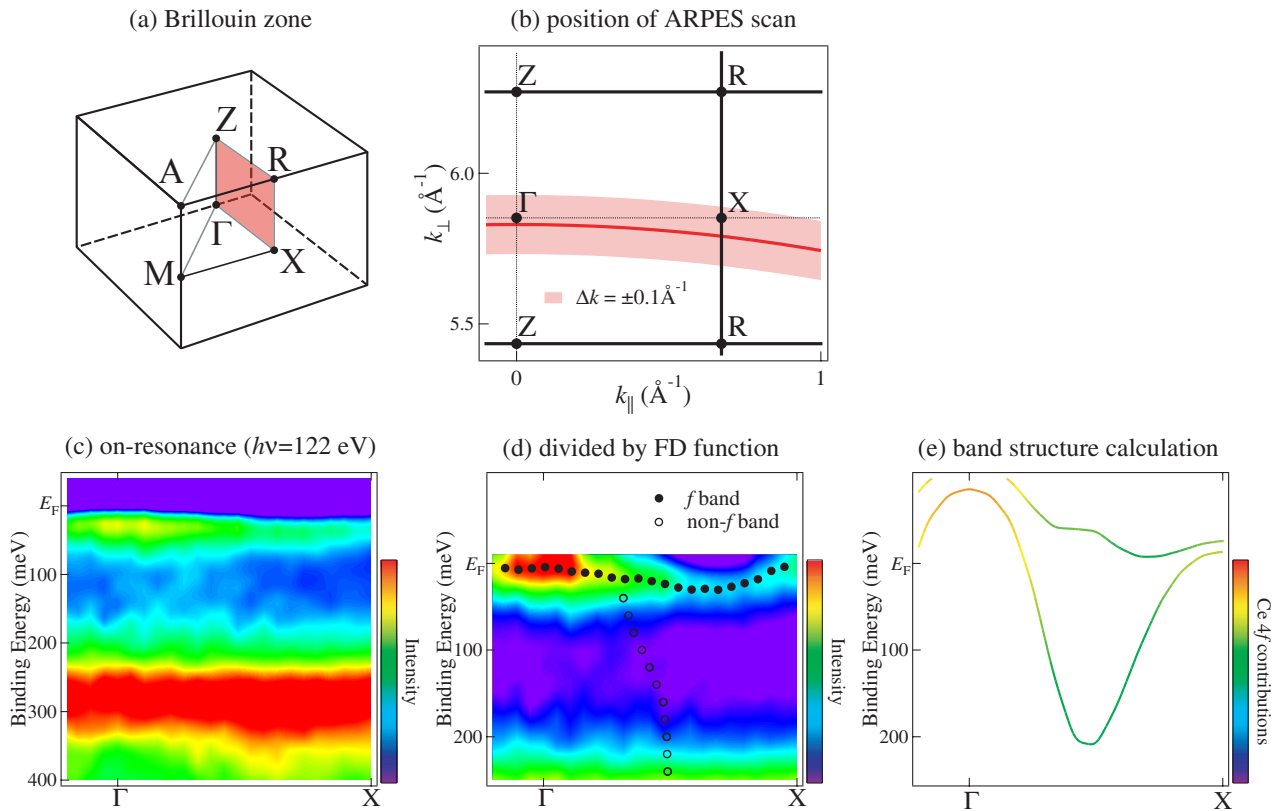

FIG. 3: (color) ARPES spectra of CeIrIn$_5$. (a) Brillouin zone (b)Position of ARPES cuts in $k$ space. (c)On-resonance ARPES spectra (d) On-resonance divided by the Fermi-Dirac function. (e)Result of the band structure calculation along the Γ-X direction, together with the positions of peaks in the on-resonance spectra and the He I spectra.

Figure 3(e) shows energy band dispersions calculated using local density approximation (LDA) for the Γ-X line. The degree of the Ce 4$f$ contribution is also indicated by the color of each line. At first glance, the calculated overall band dispersion seems very different from the observed dispersions of quasi-particle bands. However, the shape of the Ce 4$f$ dispersion seems to correspond to the band around $E_F$ but being extremely narrowed. Although the Fermi surface crossing points are different, this Ce 4$f$ dispersion can be interpreted as the band around $E_F$ being extremely narrowed. The calculated energy dispersion of this band is about 300 meV, which is about 10 times larger than the experimentally observed Ce 4$f$ dispersion, and the size of Fermi surface is smaller in the experiment than in the calculation. The calculated band mass is estimated to be $m_b = 0.64m_0$. In the dHvA study of CeIrIn$_5$, a small spherical FS having $m^* = 6.3m_0$ has observed, and it was suggested that this branch is originated from this calculated band [12]. This electron mass is ten times heavier than that of the calculation, consistent with our observation of the narrowing of the Ce 4$f$ band dispersion. In the dHvA experiments, the size of the observed Fermi surface is about 40% of the calculation. In the present experiment, although it is difficult to estimate the exact size of the Fermi surface, its size is smaller than the calculation, and the tendency seems to be consistent with the dHvA data. According to the calculation, this Fermi surface has very large contributions from Ce $4f$ states, and therefore the correlation effects are taking an important role.

Accordingly, we have found that although the global features of the angle-integrated spectra of CeIrIn$_5$ can be understood within the framework of the impurity model [13], the contribution from the small itinerant component is observed as the dispersive Ce $4f$ band near $E_F$, and explains the observation of the heavy quasi-particle bands in CeIrIn$_5$ in the dHvA experiments [12]. This dual nature of $f$ states observed in CeIrIn$_5$ is originated from the very small but substantial hybridization to produce the quasi-particle band. When the hybridization is smaller than CeIrIn$_5$, the quasi-particle band will be vanished as is observed in CeRhIn$_5$. On the other hand, if the hybridization is larger than CeIrIn$_5$, the localized nature will disappear. In fact, the RPES spectra of CeCoIn$_5$[23], which has the largest hybridization strength among the series of compounds, is very different from those of CeIrIn$_5$ as well as from CeRhIn$_5$, suggesting that $f$ states are not no longer "nearly localized" in more strongly hybridized compounds. Therefore, the strength of the hybridization in CeIrIn$_5$ is very close to the critical value for the formation of the quasi-particle band, and is the origin of the dual nature of $f$ states. Pagliuso et al. [7] have inferred that the "band like" nature of the $f$ states is important for the coexistence of magnetism and superconductivity for Ce-based compounds. However, we have found the nearly localized but slightly delocalized nature of $4f$ states even for CeIrIn$_5$, which is a non-magnetic superconductor. Therefore, the nearly localized nature with small itinerant component may be essential for such coexistence of magnetism and superconductivity in Ce-based compounds located near QCP.

## IV. CONCLUSION

In conclusion, we have found that although the Ce $4f$ electrons in CeIrIn$_5$ are nearly localized, there exists the small itinerant contribution observed as the quasi-particle band near $E_F$. This quasi-particle band has an energy dispersion of about 30 meV, consistent with the results of dHvA and optical conductivity measurements, and is responsible for the superconductivity. This dual nature of $f$ states is originated from the very small but substantial hybridization to produce the quasi-particle band in CeIrIn$_5$.

### Acknowledgments

We acknowledge the useful discussion with A. Chainani and M. Taguchi of RIKEN. The synchrotron radiation experiments have been done under the approval of HSRC (Proposal No. A01-27).